\documentclass[twoside,slac_one]{revtex4}
\usepackage{graphicx}
\usepackage{fancyhdr}
\usepackage{amsmath} 
\usepackage{bm}
\usepackage{amsxtra}
\usepackage{amssymb}
\usepackage{amsthm}
\usepackage{latexsym}
\usepackage{lscape}
\usepackage{subfigure}
\usepackage{upgreek}

\newcommand{\Kshort}{\ensuremath{K^{0}_{\mathrm{s}} }{} }

\begin{document}

\title{Performance of Particle Identification with the ATLAS Transition Radiation Tracker}

%

\author{Elizabeth Hines on behalf of the ATLAS collaboration}
\affiliation{Department of Physics and Astronomy, University of Pennsylvania, Philadelphia, PA, USA}

\begin{abstract}
The ATLAS Transition Radiation Tracker (TRT) is the outermost of the
three sub-systems of the ATLAS Inner Detector at the Large Hadron
Collider at CERN. In addition to its tracking capabilities, the TRT
provides particle identification (PID) ability through the detection
of transition radiation X-ray photons.  The latter functionality
provides substantial discriminating power between electrons and
hadrons in the momentum range from 1 to 200 GeV. In addition, the
measurement of an enhancement of signal time length, which is related
to high specific energy deposition (dE/dx), can be used to identify
highly ionizing particles, increasing the electron identification
capabilities at low momentum and improving the sensitivity of searches
for new physics.  This talk presents the commissioning of TRT PID
during early 2010 7 TeV data taking.  Performance in 2010 and 2011
demonstrating the TRT's ability to identify electrons, complementary
to calorimeter based identification methods, will also be shown.

\end{abstract}

\maketitle

\thispagestyle{fancy}


\section{Introduction}

The ATLAS Inner Detector (ID) is composed of three detector
sub-systems: the silicon-based Pixel and SemiConductor Tracker (SCT)
detectors, and the gaseous drift tube Transition Radiation Tracker
(TRT) \cite{DetectorPaper}. The TRT is the outermost of the three
sub-systems. It employs a unique design which combines tracking
measurements with particle identification based on detection of
transition radiation (TR).  The detection of TR allows for
discrimination between electrons and pions over the energy range 1-200
GeV and is a crucial component of the electron selection criteria in
ATLAS \cite{ElectronPaper}. These proceedings present the particle
identification (PID) performance of the TRT observed in $\sqrt{s} = 7$
TeV proton-proton collision data collected with the ATLAS detector at
the Large Hadron Collider (LHC) in 2010.

Transition radiation is emitted when a highly-relativistic charged
particle with a Lorentz factor $\gamma \gtrsim 10^{3}$ traverses
boundaries between materials of differing dielectric constants. The
active region of the TRT detector contains almost 300,000 straw drift
tubes of 4mm diameter. The space between the straws is filled with
radiator material. The TR photons (soft X-rays) emitted in the
radiator are absorbed in the active gas inside the straw tubes, which
serve as detecting elements both for tracking and for particle
identification.

Particle identification in the TRT can be further augmented at momenta
p $\lesssim$ 10 GeV through measurements of the time-over-threshold
(ToT) of the straw signals, which vary as a function of energy
deposition (dE/dx) in the straws. To achieve the best electron-pion
separation, TR and dE/dx-based measurements are combined in a single
likelihood function for a particle type.

\section{Transition Radiation Tracker}

The TRT is a straw tracker composed of 298,304 carbon-fiber reinforced
Kapton straws, arranged in a barrel and two symmetrical end-cap
configurations \cite{StrawPaper}. The barrel section covers $560 < R <
1080$ mm and $|z| < 720$ mm and has the straws aligned with the
direction of the beam axis \cite{BarrelPaper}.\footnote{ATLAS uses a
  right-handed coordinate system with its origin at the nominal
  interaction point (IP) in the center of the detector and the z-axis
  coinciding with the axis of the beam pipe. The x-axis points from
  the IP to the center of the LHC ring, and the y-axis points
  upward. Cylindrical coordinates (R, $\phi$) are used in the
  transverse plane,$\phi$ being the azimuthal angle around the beam
  pipe and R, the distance from the IP in the radial direction. The
  track pseudo-rapidity is defined as $\eta = - \ln(\theta/2)$, where
  the polar angle $\theta$ is the angle between the track direction
  and the z axis} The two end-cap sections cover $827 < |z| < 2744$ mm
and $617 < R < 1106$ mm and have the straws arranged in planes
composing wheels, aligned perpendicular to the beam axis, pointing
outwards in the radial direction \cite{EndcapPaper}. The TRT extends
to pseudo-rapidity $|\eta| = 2$. The average number of TRT hits per
track is around 34, except in the transition region between barrel and
end-caps and at the edge of the acceptance ($|\eta| \lesssim 1.7$)
where it is reduced to approximately 25 hits. Polypropylene fibers
interwoven between straw layers are used in the barrel for radiator
material and regular polypropylene foils in the end-caps. The straws
are filled with a gas mixture of 70\% Xe, 27\% CO$_2$ and 3\%
O$_2$. Xenon was chosen for its high efficiency to absorb TR photons
of typical energy 6$-$15 keV.

The TRT operates as a drift chamber: when a charged particle traverses
the straw, it ionizes the gas, creating about 5-6 primary ionization
clusters per mm of path length. The straw wall is held at a potential
of about $-$1530V with respect to a 31 $\mu$m diameter gold-plated
tungsten wire at the center that is referenced to ground. The
electrons drift toward the wire and cascade in the strong electric
field, producing a detectable signal. On each wire the signal is
amplified, shaped and discriminated against two adjustable thresholds,
a low-threshold (LT) of about 300 eV and a high-threshold (HT) of
about 6-7 keV \cite{ElectronicsPaper}. The two thresholds allow for
simultaneous measurement of tracking information and identification of
characteristic large energy deposits due to the absorption of TR
photons. For any triggered event, the TRT reads out data over three
nominal bunch crossing periods, 3x25 ns.\footnote{During 2010 running
  used in these proceedings, the spacing between bunches was 150 ns or
  greater.} The measured drift times are at most $\sim$ 50 ns. Low
threshold information is read-out separately in time intervals of 3.12
ns length so that each bunch-crossing is divided into eight time
bins. The first low threshold 0 $\rightarrow$ 1 transition marks the
leading edge (LE) of the signal (hit), and the leading edge time
t$_{LE}$ is defined as the center of the first bin set to
1. Similarly, the last 1 $\rightarrow$ 0 transition is called the
trailing edge (TE) of the hit. High-threshold information is recorded
at a coarser granularity, every 25 ns (once per bunch-crossing),
giving three HT bits per hit for a triggered event. A hit is said to
be a HT hit if any of the three HT bits is high.  The leading
(trailing) edge time depends on the time when the closest (furthest)
ionization electron cluster arrives at the wire at the center of the
straw. The leading edge time is thus directly related to the
track-to-wire distance r$_{track}$. If the furthest electrons were
always produced exactly at the straw wall and drifted for the full
straw radius of 2mm, t$_{TE}$ time would be independent of
r$_{track}$. Due to the finite interaction length (and thus the
limited number of primary ionization clusters) and signal shaping
effects, this is not always the case. A particle that deposits more
energy will, on average, have a higher signal, exceed the threshold
sooner, and fall back below threshold later.  Thus, larger energy
deposits result in an earlier LE, later TE and longer ToT on
average. This correlation can be used to obtain a ToT-based dE/dx
estimate.

\section{Data samples and trigger requirements}

Data from proton-proton collisions at the LHC at $\sqrt{s}$ = 7 TeV
recorded by the ATLAS detector in 2010 were used for the studies
reported in these proceedings. The detector response for electrons was
studied with samples of reconstructed photon conversions and Z boson
decays, in order to explore two different momentum ranges and exploit
the abundance of photon conversions in early data.  The detector
response to pions was studied using the same minimum-bias data set as
for photon conversions.

A minimum bias trigger was used to record the data set used for the
reconstruction of photon conversions and pion candidates. During the
initial low-luminosity running period from April 15 to June 5, 2010,
the events were collected in the minimum bias trigger stream at a rate
that was typically between 40Hz and 200Hz, providing a high statistics
sample of electrons from photon conversions. This data set corresponds
to an integrated luminosity of approximately $\mathcal{L}$ = 9 nb$^{-1}$.

Data recorded June 24, 2010 - October 29, 2010, corresponding
to an integrated luminosity of $\mathcal{L}$ = 35 pb$^{-1}$, was used to
reconstruct electron candidates from Z boson decays. Events were
required to be triggered by an electron trigger that has close to
100\% efficiency for electrons from Z boson decays selected in this
analysis.

The LHC bunch spacing during both running periods was 150 ns or
greater. Pile-up from multiple interactions per bunch crossing was
small. The average number of minimum bias interactions per beam
crossing was less than 0.2 in the data set used for photon
conversions, and about three in the data set used to reconstruct the
sample of Z bosons.

The results observed in data were compared to Monte Carlo (MC)
simulations \cite{SimulationPaper}. The detector response to electrons
from photon conversions and pions in data were compared to {\tt
  Pythia} non-diffractive minimum bias MC simulation. The electrons
from Z boson decays were compared to {\tt Pythia} $Z \to e^{+}e^{-}$
MC simulation.

\subsection{Electron candidates}

Photon conversions to electron-positron pairs were used to reconstruct
a pure sample of electrons in early data. The photon conversion
candidates \cite{ConvConfNote} are required to have two tracks, each
with a minimum of 20 TRT hits and four silicon (SCT and Pixel)
hits. The conversion vertex is required to be well reconstructed and
to be at least 60 mm away from the primary vertex in the radial
direction. To improve the sample purity, a tag and probe method is
applied to the two tracks of the selected photon conversion
candidates. The tag leg is required to have a ratio of the number of
TRT high-threshold hits to total TRT hits of at least 0.12, which
corresponds to at least three high-threshold hits on a track with the
minimum total number of 20 TRT hits. For a conversion candidate
passing these requirements, the probe leg is declared to be an
electron candidate. The two tracks are treated independently; if both
of the tracks pass the tag requirement, each is also used as a
probe. Over 500,000 electron candidates satisfy these selection
criteria, providing a high statistics sample of electron candidates at
the early stages of collision data-taking.

A second sample of electron candidates is obtained from the
reconstruction of $Z \to e^{+}e^{-}$ decays. Electrons from this
sample have higher momenta, and can thus be used to probe the TR
performance at higher values of $\gamma$. Electron candidates are
required to pass the calorimeter based ``medium'' electron selection
criteria \cite{ElectronPaper}, and to have an innermost Pixel layer
(b-layer) hit.  Candidate events are required to have two such
electrons, with a reconstructed di-lepton invariant mass in the range
75 $-$ 105 GeV, based on measurements from their calorimeter
clusters. Electrons from Z boson decays are treated in the same way as
those from photon conversions. The tag leg is required to have a TRT
high-threshold ratio greater than 0.12, and both legs are required to
have at least 20 TRT hits.

\subsection{Pion candidates}

Pion candidates are selected from reconstructed particle tracks that
have a minimum of 20 TRT and four silicon hits. Further selection
criteria are applied to reject electrons, protons and kaons. Any track
that does not have a hit in the innermost Pixel layer or that is
reconstructed as a part of a photon conversion candidate is excluded.
These two requirements reduce electron contamination from photon
conversions, which is the dominant source of electrons in the minimum
bias data. In addition, any track with a measured dE/dx above 1.6
MeVg$^{-1}$cm$^2$ in the Pixel detector is excluded in order to reduce
the contamination from protons (and to a lesser extent kaons) at low
momentum. A track passing these requirements is declared to be a
pion candidate. 

\section{Transition radiation and high-threshold hits}

\begin{figure}[ht]
  \centering
  \subfigure[ TRT barrel]{\includegraphics[width=80mm]{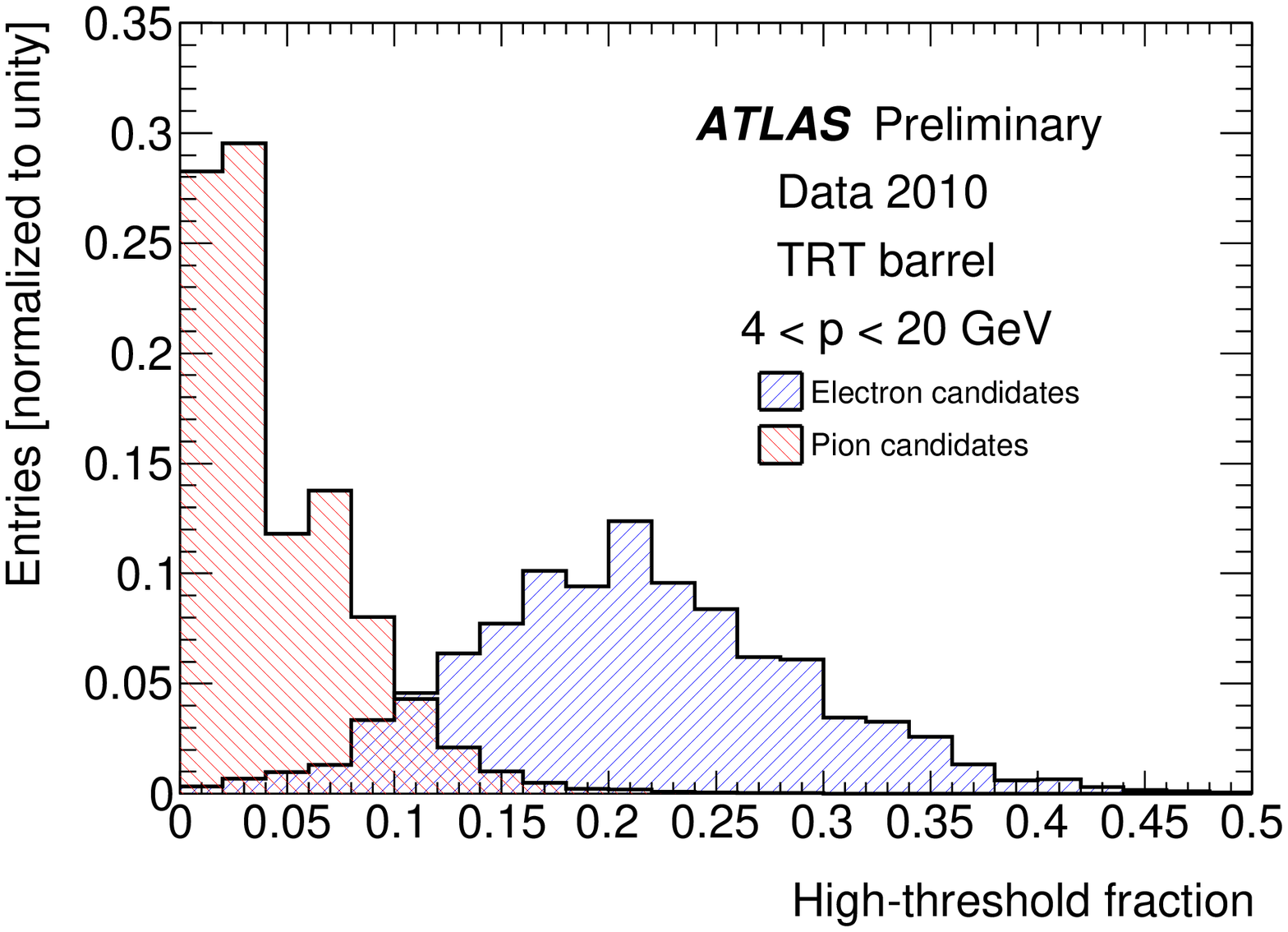}
    \label{fig:SeparationBarrel}}
  \subfigure[ TRT end-caps]{\includegraphics[width=80mm]{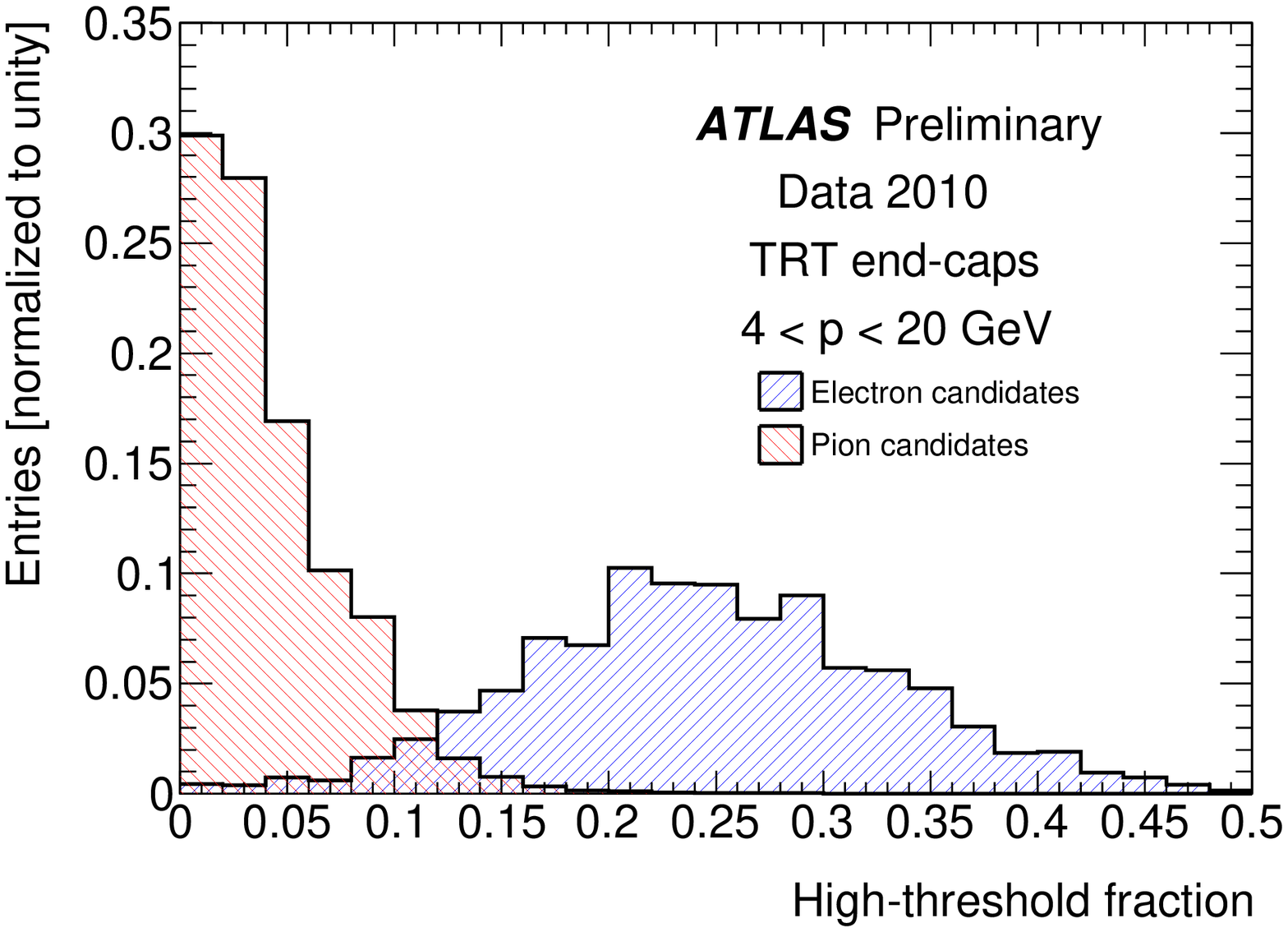}
    \label{fig:SeparationEndcap}}
  \caption{ The HT fraction for electron and pion candidates} 
  \label{fig:Separation}
\end{figure}

This section presents the results of HT studies in electron from
photon conversions and pion samples. Figure \ref{fig:Separation} shows
the HT fraction distributions for electron and pion candidates. The HT
fraction is defined as the ratio hits on track that exceed the high
threshold to the total number of hits on track. The distribution for
electrons is shown in Fig. \ref{fig:Separation} is clearly shifted to
higher values. This value is defined on a per track basis, whereas the
high-threshold probability is defined as the total number of
high-threshold hits summed over all candidates divided by the total
number of hits summed over all candidates.  The following sections
show the HT dependence on the $\gamma$ factor, and performance of a
requirement on the HT fraction, in terms of the electron efficiency
and pion misidentification probability. Finally, validation of
hardware settings with 7 TeV collision data are presented.

\subsection{Transition radiation onset}

\begin{figure}[ht]
  \centering
  \subfigure[ Barrel]{\includegraphics[width=80mm]{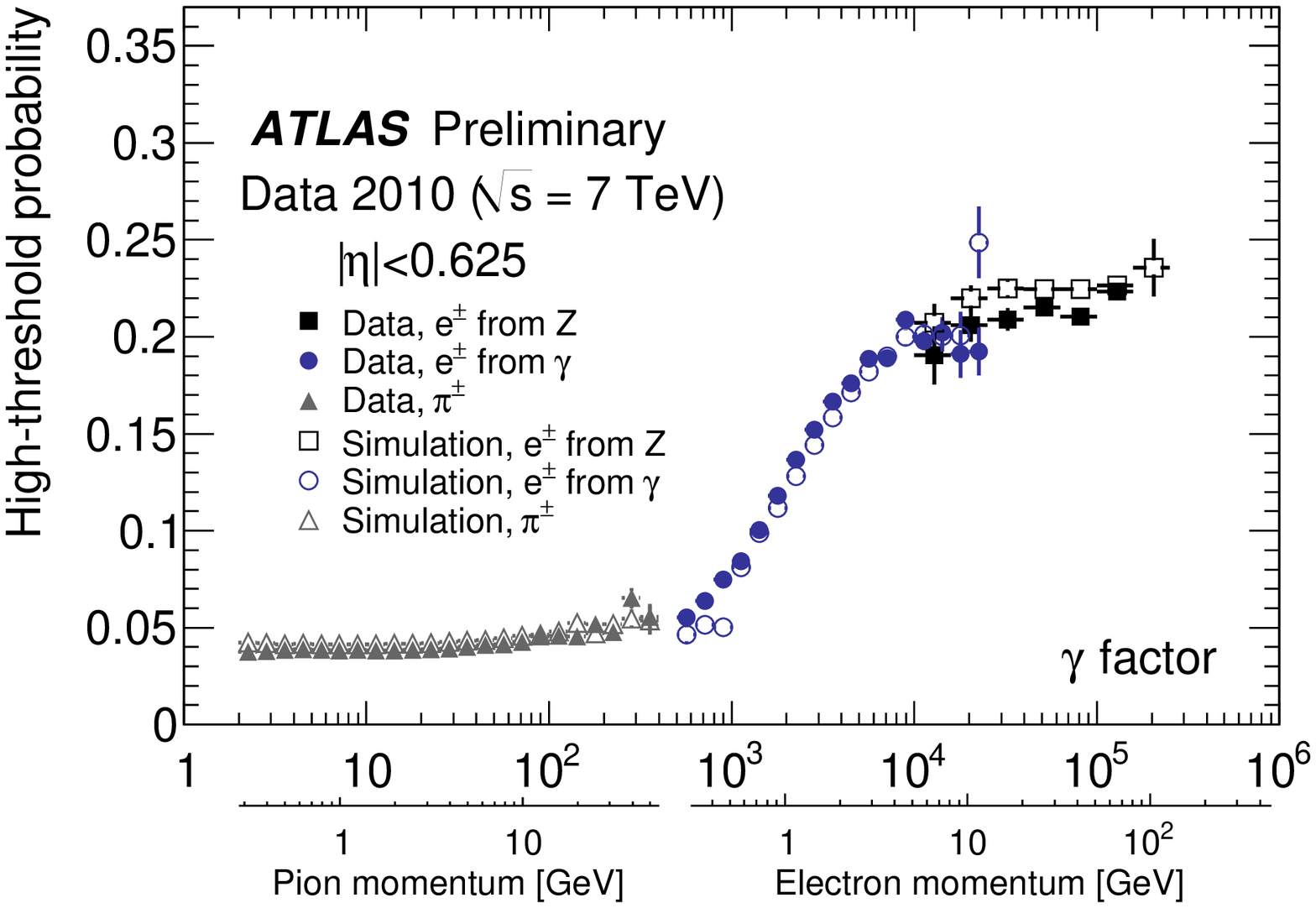}
    \label{fig:TurnOnCurveBarrel}}
  \subfigure[ End-cap A-type wheels]{\includegraphics[width=80mm]{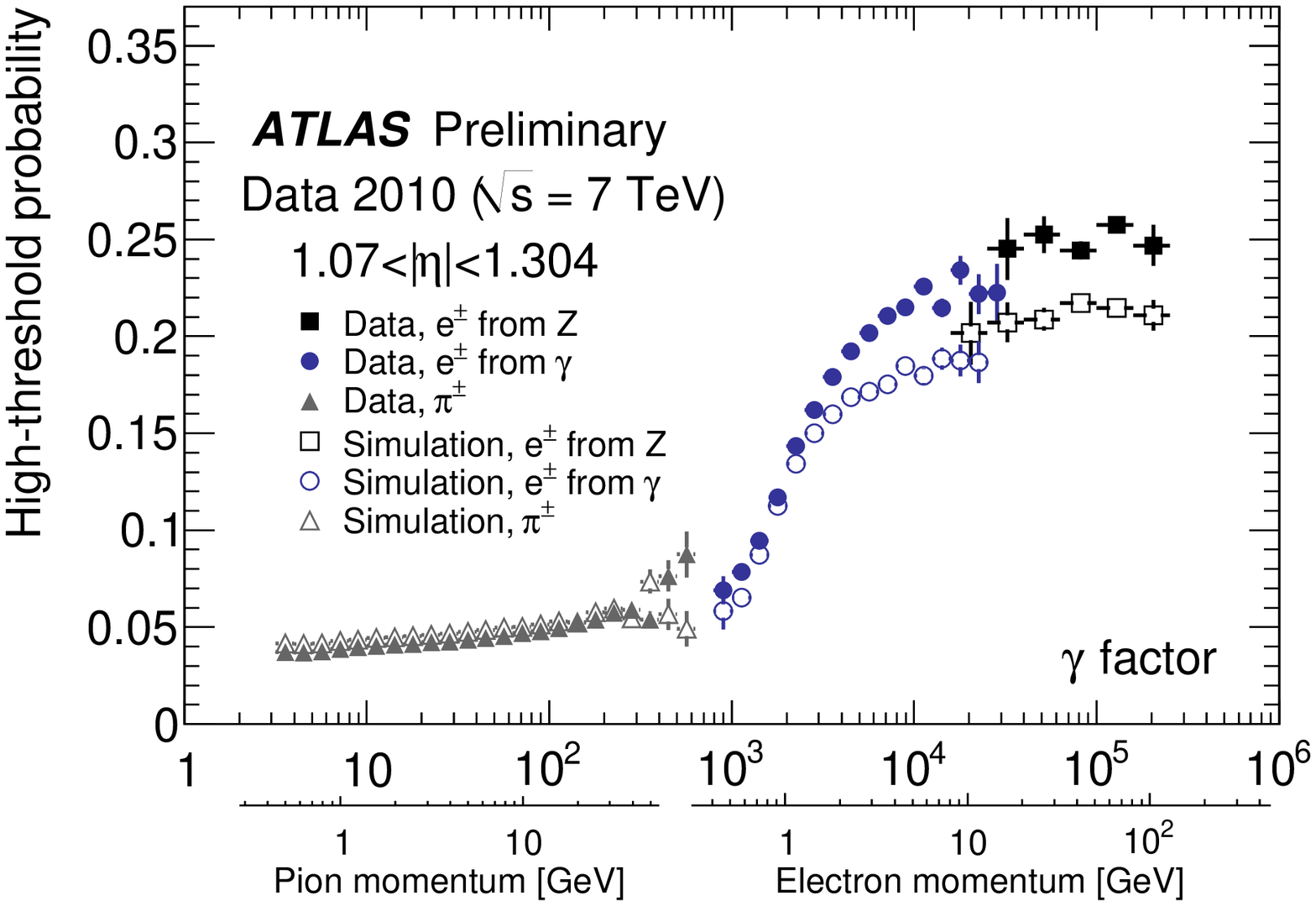}
    \label{fig:TurnOnCurveEndcapA}}
  \subfigure[ End-cap B-type wheels]{\includegraphics[width=80mm]{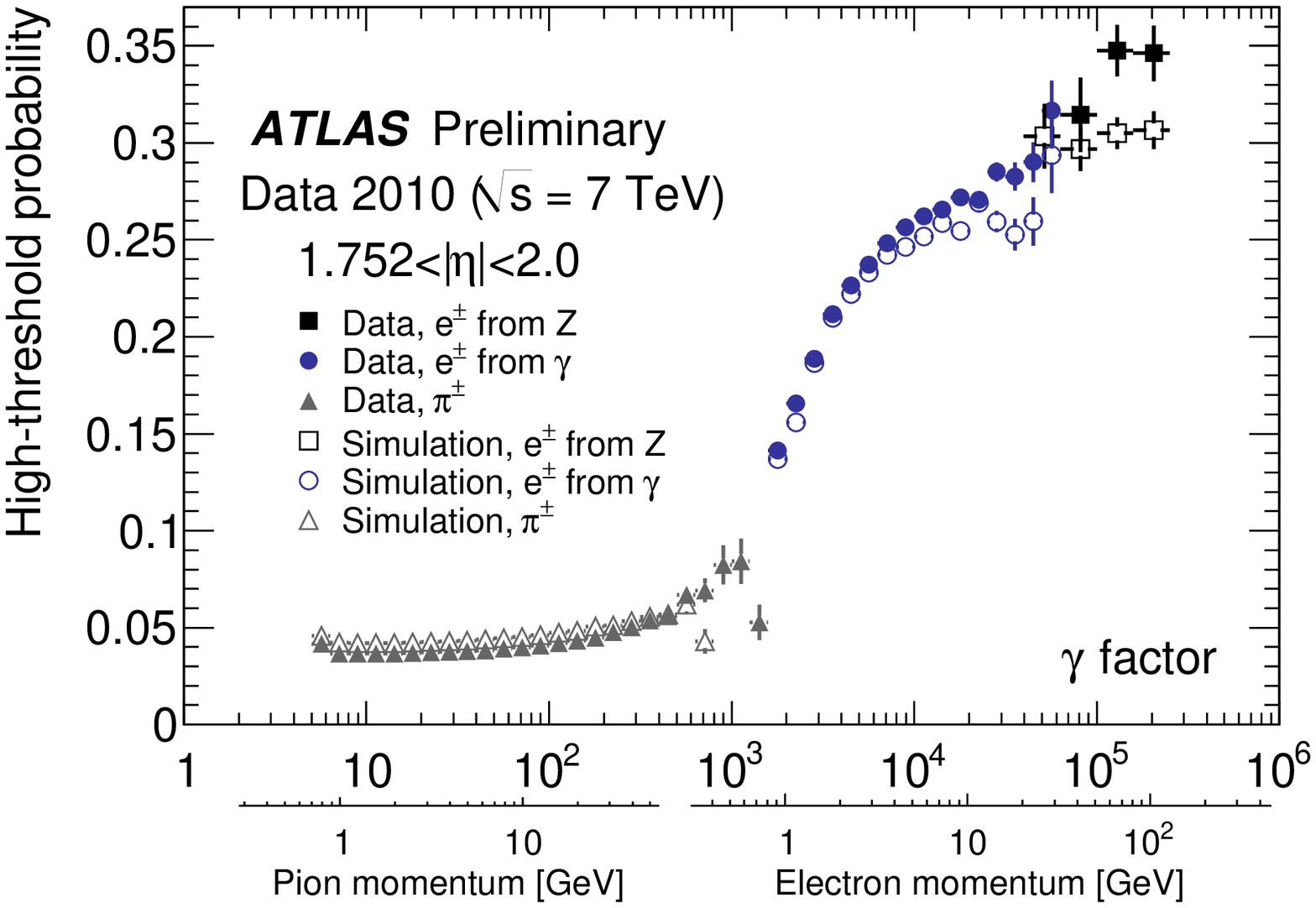}
    \label{fig:TurnOnCurveEndcapB}}
  \caption{The high-threshold turn on curve, separated into detector
    regions by track $\eta$}
  \label{fig:TurnOnCurves}
\end{figure}

The first step towards establishing electron identification with the
TRT is to observe the expected increase in the average number of HT
hits with $\gamma$. The increase has been observed in 2004 test-beam
data [13],cosmic-ray data [14] and for collision data at $\sqrt{s} =
900$ GeV [15].  The HT probability observed in 7 TeV collision data is
shown in Fig. \ref{fig:TurnOnCurves} and is consistent with earlier
measurements.  The results are shown separately for five intervals in
pseudo-rapidity $\eta$ reflecting different detector regions. The
errors shown are statistical only. The average HT fraction was
evaluated for tracks in bins of the Lorentz factor $\gamma$. The
pions, electrons from photon conversions and electrons from Z boson
decays cover different $\gamma$ ranges. For the electron candidates,
the sharp turn-on of the transition radiation can be seen, with the HT
probability increasing rapidly from 0.05 to a plateau of 0.2 − 0.3
depending on $\eta$ region. The HT plateau level in the end-cap region
is higher than in the barrel.  Electrons from the reconstructed Z
decays allow studies of HT probability at $\gamma \sim 10^5$, which
can not be accessed with electrons from photon conversions. Small
differences in the HT probability for the electrons from conversions
and $Z \rightarrow ee$ decays in the overlapping $\gamma$ range can
not be resolved at the current statistical uncertainty.

The pion candidates shown in Fig. \ref{fig:TurnOnCurves} populate the
region $\gamma < 10^3$. In this $\gamma$ range, HT hits are caused by
large ionization energy deposits due to Landau dE/dx fluctuations. The
HT probability for pion candidates increases gradually from about 0.04
at $\gamma \sim 1$ to about 0.07 at $\gamma \sim$ 700 (p $\sim$ 100 GeV)
due to the rise of $<dE/dx>$ with increasing track momentum. This
behavior was cross-checked with a sample of pion candidates from
\Kshort decays that has higher pion purity, and the results were in
good agreement.

\subsection{Electron efficiency and pion misidentification probability}

\begin{figure}[ht]
  \centering
  \subfigure[]{\includegraphics[width=80mm]{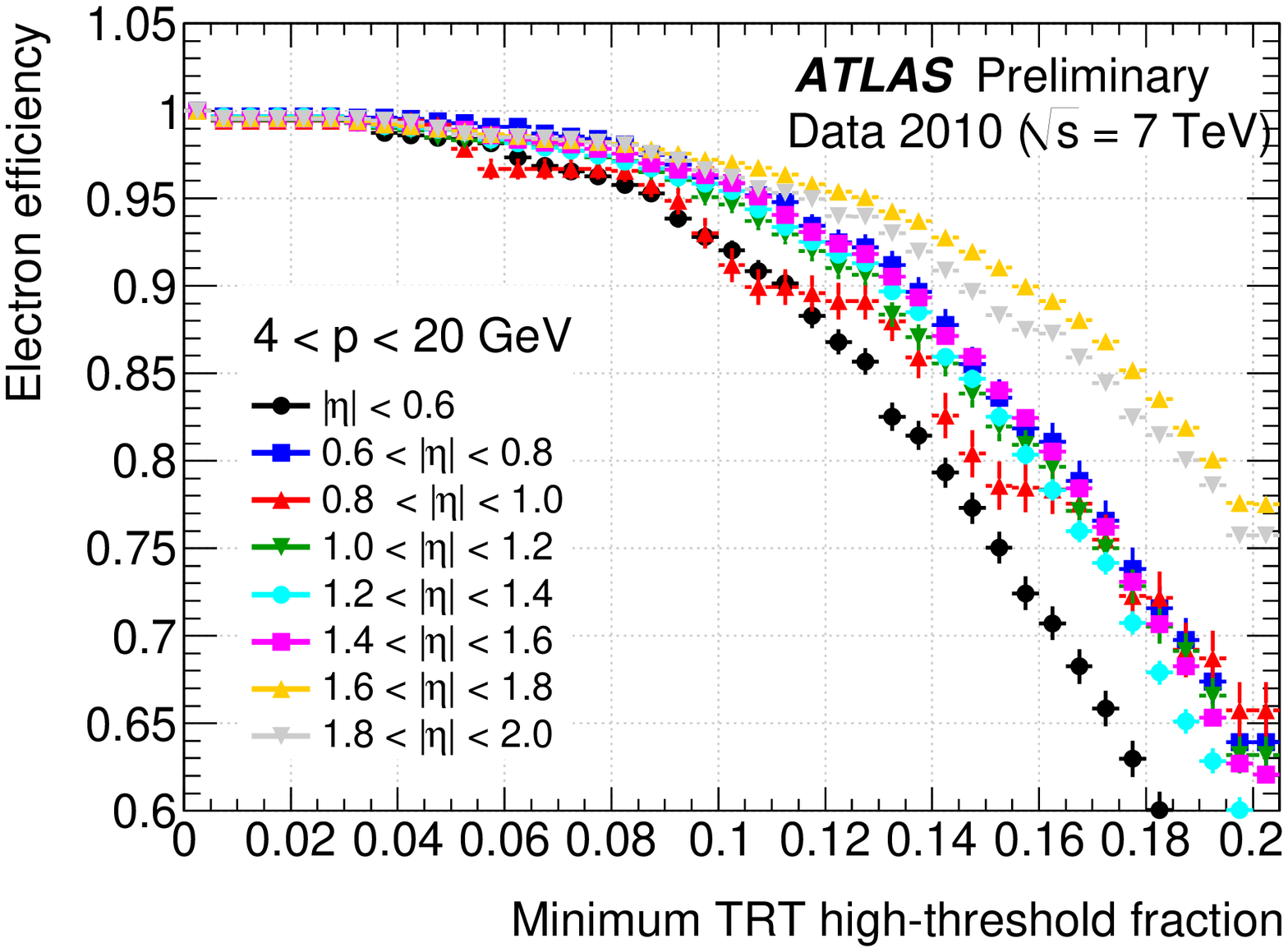}
    \label{fig:EffVsHTEl}}
  \subfigure[]{\includegraphics[width=80mm]{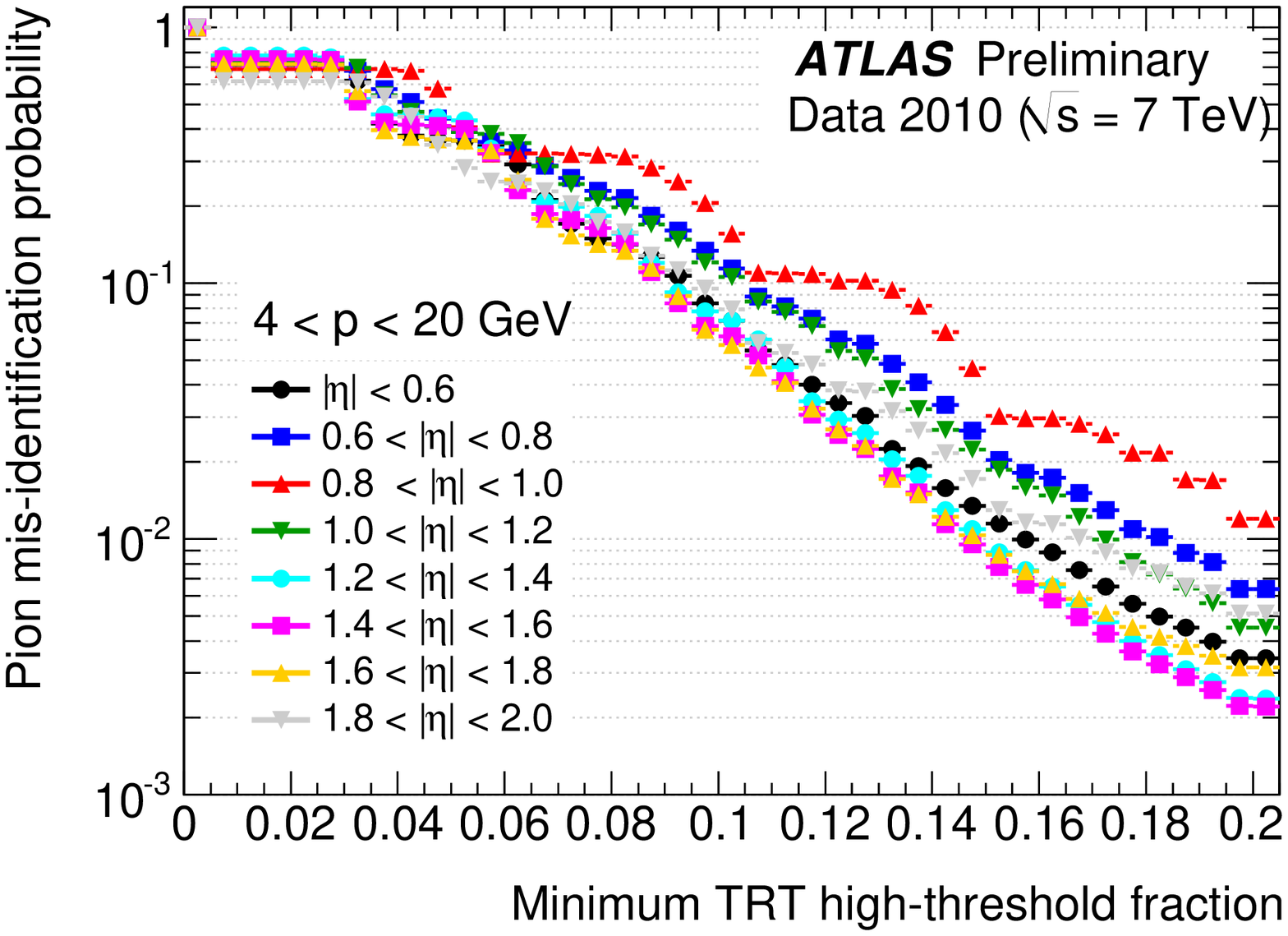}
    \label{fig:EffVsHTPion}}
  \caption{The fraction of electron (a) and pion (b) candidates that
    pass a cut on high-threshold fraction.}
  \label{fig:EffVsHT}
\end{figure}

\begin{figure}[ht]
  \centering
  \includegraphics[width=80mm]{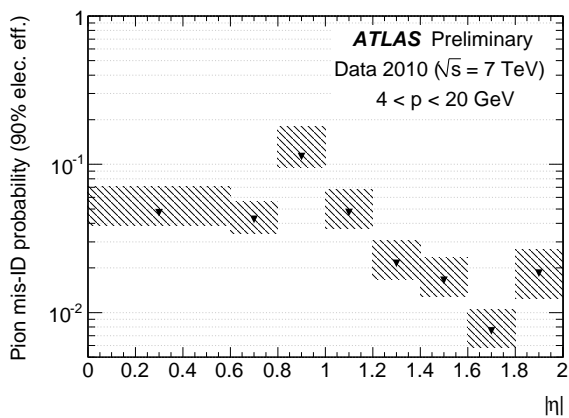}
  \caption{Pion mis-ID probability for the HT faction criteria that
    gives 90\% electron efficiency, }
  \label{fig:PionRej}
\end{figure}

The HT-based electron-pion separation demonstrated in
Fig. \ref{fig:Separation} is utilized by a requirement of a minimum HT
fraction for electron identification. Figure \ref{fig:EffVsHTEl} shows
the fraction of electron candidates that pass a HT fraction selection
requirement, in bins of $|\eta|$.  The pion misidentification
probability p$_{\pi\rightarrow e}$ is the probability for a pion to
pass an electron HT fraction selection criteria and is shown in
Fig. \ref{fig:EffVsHTPion}. The pion rejection power is
$1/p_{\pi\rightarrow e}$. A direct comparison of the electron
efficiency and pion misidentification probability is shown in Fig.
\ref{fig:PionRej}.  A benchmark point of a cut on high-threshold
fraction that has a 90\% electron efficiency is used.  The uncertainty
on the pion misidentification probability shown in
Fig. \ref{fig:PionRej} was estimated by varying the selection criteria
such that the electron efficiency changed by $\pm$2\%. The range of
$\pm$2\% is sufficiently big to include the uncertainties due to
hadron contamination in the electron sample of about 1\%.

The minimum HT fraction that is required for an electron to pass the
ATLAS ``tight'' electron selection requirement \cite{ElectronPaper},
the corresponding efficiency for electrons to pass this criterion as
well as the pion misidentification probability are summarized in Table
\ref{tab:tightEff}. The current HT fraction selection criteria were
determined based on MC studies prior to the start of collision
data-taking, and was chosen such that a pion rejection factor of at
least 10 would be achieved after applying the HT fraction electron
selection criteria \cite{DetectorPaper}. In the range 0.625 $< |\eta|
< $1.07, only a factor of four was achieved due to fewer hits on track
in the transition region and a relatively large HT hit probability for
pions for geometric reasons. In the highest $\eta$ bin, the pion
rejection factor is almost 100.

\begin{table}[ht]
\begin{center}
\caption{Fraction of electron and pion candidates that pass the HT
  fraction cut used in ``tight'' electron identification for each
  $\eta$ range.  Errors given are statistical only.}
\begin{tabular}{|l|c|c|c|}
\hline \textbf{$\eta$ Range} & \textbf{Minimum HT fraction} & \textbf{Electron Efficiency} &
\textbf{Pion misidentification probability}
\\
\hline $0.0 \rightarrow 0.625$   & 0.085 & 0.953 $\pm$ 0.004 & 0.1268 $\pm$ 0.0003 \\
\hline $0.625 \rightarrow 1.07$  & 0.085 & 0.961 $\pm$ 0.005 & 0.2420 $\pm$ 0.0004 \\
\hline $1.07 \rightarrow 1.304$  & 0.115 & 0.921 $\pm$ 0.005 & 0.0473 $\pm$ 0.0001 \\
\hline $1.304 \rightarrow 1.752$ & 0.13  & 0.919 $\pm$ 0.002 & 0.0174 $\pm$ 0.0001 \\
\hline $1.752 \rightarrow 2.0$   & 0.155 & 0.882 $\pm$ 0.002 & 0.0109 $\pm$ 0.0001 \\
\hline
\end{tabular}
\label{tab:tightEff}
\end{center}
\end{table}

\subsection{Validation of hardware settings}

In order to determine the optimal average high threshold setting, data
corresponding to an integrated luminosity of 20 nb$^{-1}$ was taken
with different HT settings in July 2010, and the results of the pion
rejection study with different settings are reported in this
section. An electron trigger that maximized the number of
reconstructed photon conversion candidates was used to record these
data.

The value of the high threshold can be varied by changing the Digital
to Analogue Converter setting (DAC counts) on the Amplification,
Shaping, Discrimination, and Base-Line Restoration (ASDBLR) chip
\cite{ElectronicsPaper}, in steps of about 60 eV. Prior to the start
of collision data-taking, the average HT was adjusted to the setting
that gave the best performance at the test beam. Results from
electronics noise scans were used to correct for the large variations
in response due to variations in ground offsets. 

Validation of the overall average setting for the full detector is
reported here. To validate the average HT setting, data were recorded
with six different HT settings: nominal settings, $\pm$15 DAC counts from
nominal, $\pm$25 DAC counts from nominal, and $+$8 DAC counts from
nominal. The high-threshold settings were varied uniformly across the
entire detector.

\begin{figure}[ht]
  \centering
  \includegraphics[width=80mm]{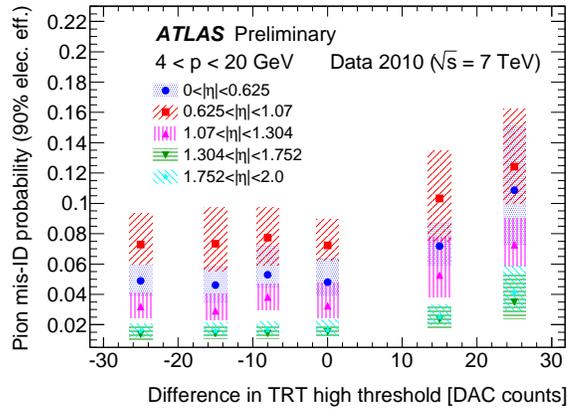}
  \caption{Pion mis-ID probability at 90\% electron efficiency as a
    function of hardware settings for different $\eta$ ranges.}
  \label{fig:SpecialRunSummary}
\end{figure}

As the threshold is decreased, the HT probability increases for both
electron and pion candidates.  The optimal average HT setting is
determined based on the pion rejection power. The HT fraction
selection criteria that gives 90\% electron efficiency was determined
for different values of high threshold settings and for different
$\eta$ bins. Figure \ref{fig:SpecialRunSummary} shows the efficiency
for a pion candidate to pass the selection criteria as a function of
the high threshold setting difference. The selection criteria at 90\%
electron efficiency was used as a reference for this study. As in the
previous section, the uncertainties were estimated by varying the
selection criteria such that the electron efficiency changed by
$\pm$2\%. For all regions, the pion misidentification probability
p$_{\pi\rightarrow e}$ is independent of the HT setting in the range
of -25 to nominal DAC count. For settings higher than nominal,
p$_{\pi\rightarrow e}$ increases. Based on these results, the
high-threshold was lowered by eight DAC counts across the detector for
2011 data-taking. The primary reason for lowering the thresholds was
to operate at stable settings, where the performance does not vary
much if the HT is slightly above or below the nominal.

\section{Time over threshold based particle identification}

The measured time over threshold is correlated with the ionization
deposit within the straw, and can thus be used to better distinguish
between electrons and pions based on their expected dE/dx. For the
purpose of the dE/dx measurement, the ToT is defined as the number of
bits above threshold in the largest single group of bits above
threshold, multiplied by the bin width. This method has a similar
performance to a method that uses all bits above threshold, and a
better performance than a method that uses t$_{TE}-$t$_{LE}$.

The ToT is subject to several systematic effects that are not related
to dE/dx. The t$_{LE}$ depends on the track-to-wire distance due to
the drift time. Due to the limited number of primary ionization
clusters, the t$_{TE}$ also depends on the track-to-wire distance. The
track-to-wire distance related variation in the measured ToT is about
10 ns. Other smaller effects that can cause variations of a few ns
along the wire length are signal attenuation (attenuation length
$\lambda$ = 4m \cite{StrawPaper}), signal reflection from the end of
the wire that is not read out, signal delay due to the propagation
along the wire and signal shaping.  These effects are taken into
account by corrections that vary with the track-to-wire distance and
distance along the straw. The track-to-wire distance dependent
corrections also take into account the dependence of the total energy
deposit within the straw on the track length.

\begin{figure}[ht]
  \centering
  \subfigure[ Barrel]{\includegraphics[width=80mm]{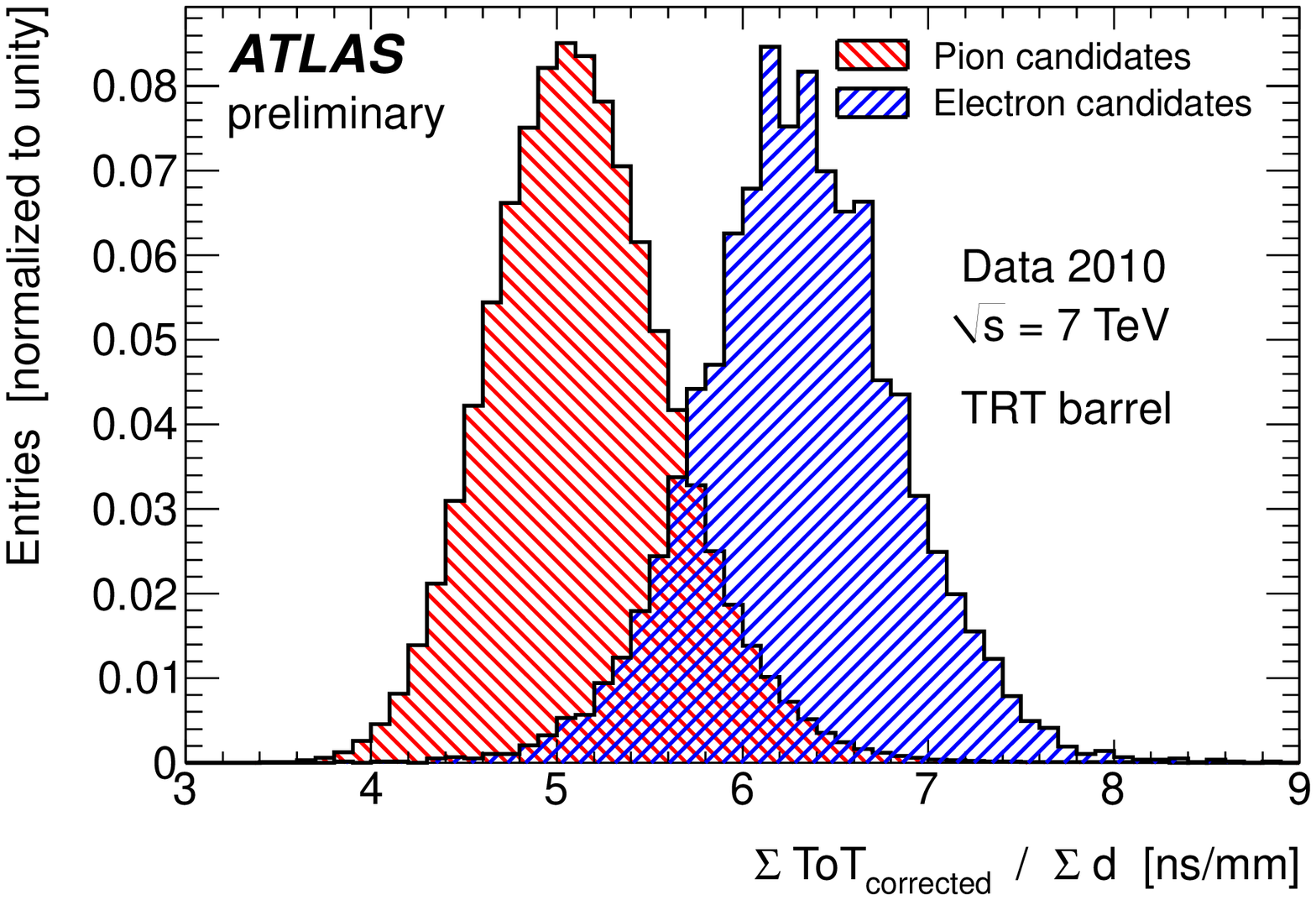}
    \label{fig:ToTSepBarrel}}
  \subfigure[ Endcap]{\includegraphics[width=80mm]{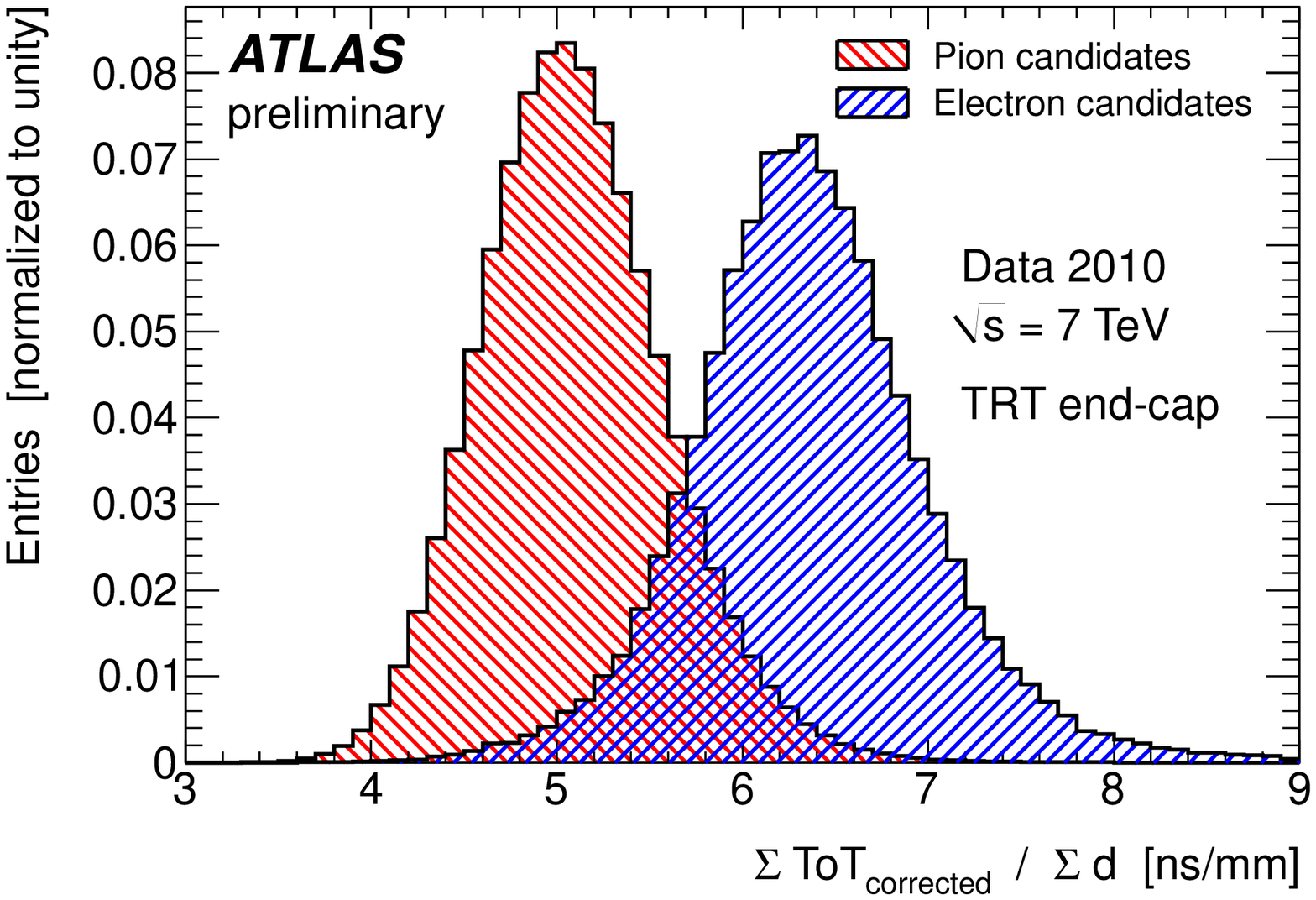}
    \label{fig:ToTSepEndcap}}
  \caption{The average time over threshold, corrected for the track
    length within the straw and distance along the straw (z in the
    barrel, R in the end-caps) }
  \label{fig:ToTSep}
\end{figure}

The ionization loss for electrons and pions differs the most for
particles of low momentum, p $<$ 10 GeV.  To achieve the best
sensitivity, all systematic effects discussed are taken into
account. Corrections are made for z dependence in the barrel and R
dependence in the end-caps. To take into account the track-to-wire
distance dependence, the average corrected ToT measurement is divided
by the average track-to-wire distance.  The track level ToT-based
discriminator is obtained by averaging corrected ToT measurements for
all hits on track that do not exceed the HT. The HT hits are not used
in order to avoid the correlation between the ToT-based variable and
the HT fraction. Figure \ref{fig:ToTSep} shows the corrected ToT
distributions for the electron and the pion candidates.

\section{Combination of HT and ToT measurements}

\begin{figure}[ht]
  \centering
  \subfigure[ Barrel]{\includegraphics[width=80mm]{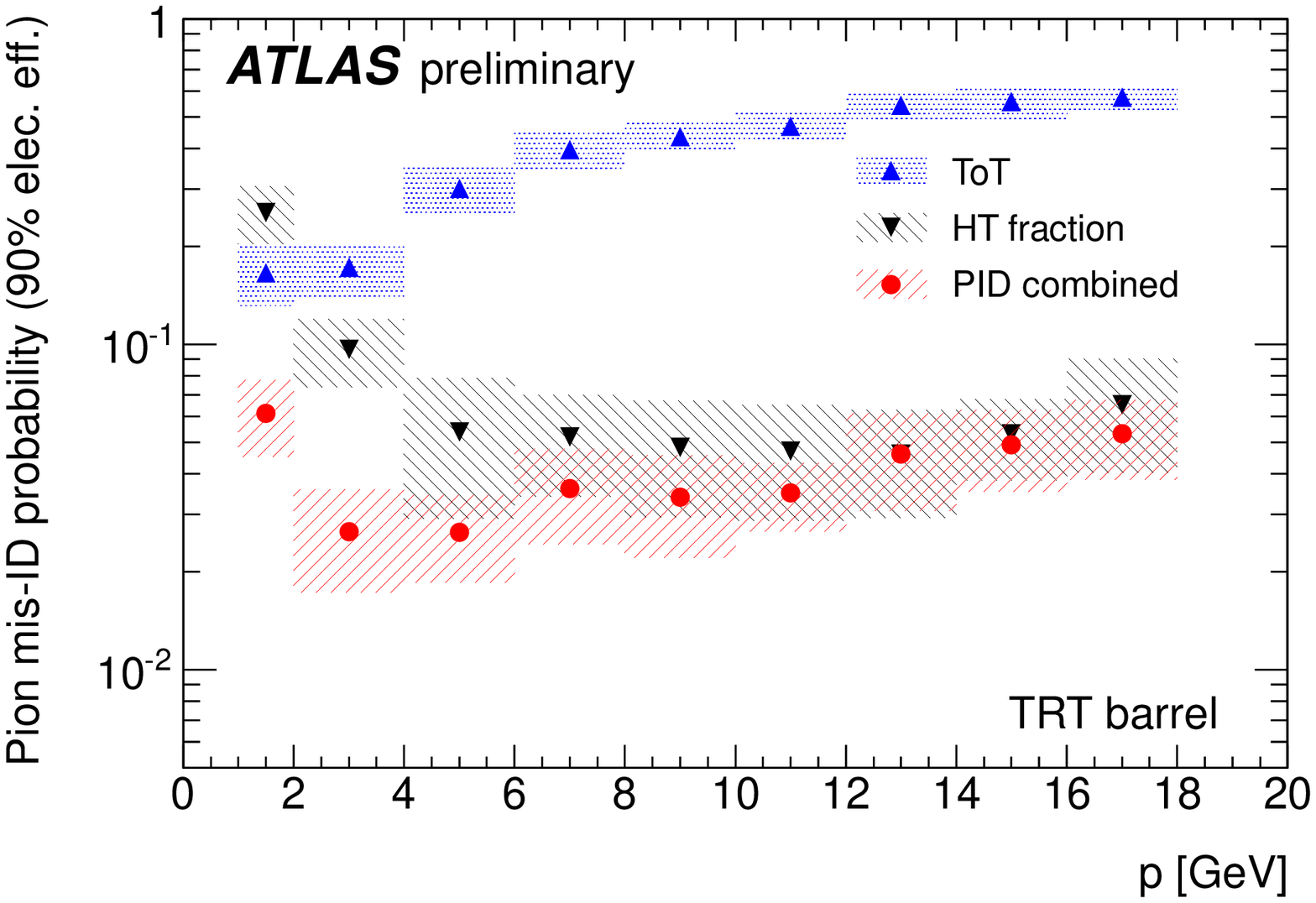}
    \label{fig:CombBarrel}}
  \subfigure[ Endcap]{\includegraphics[width=80mm]{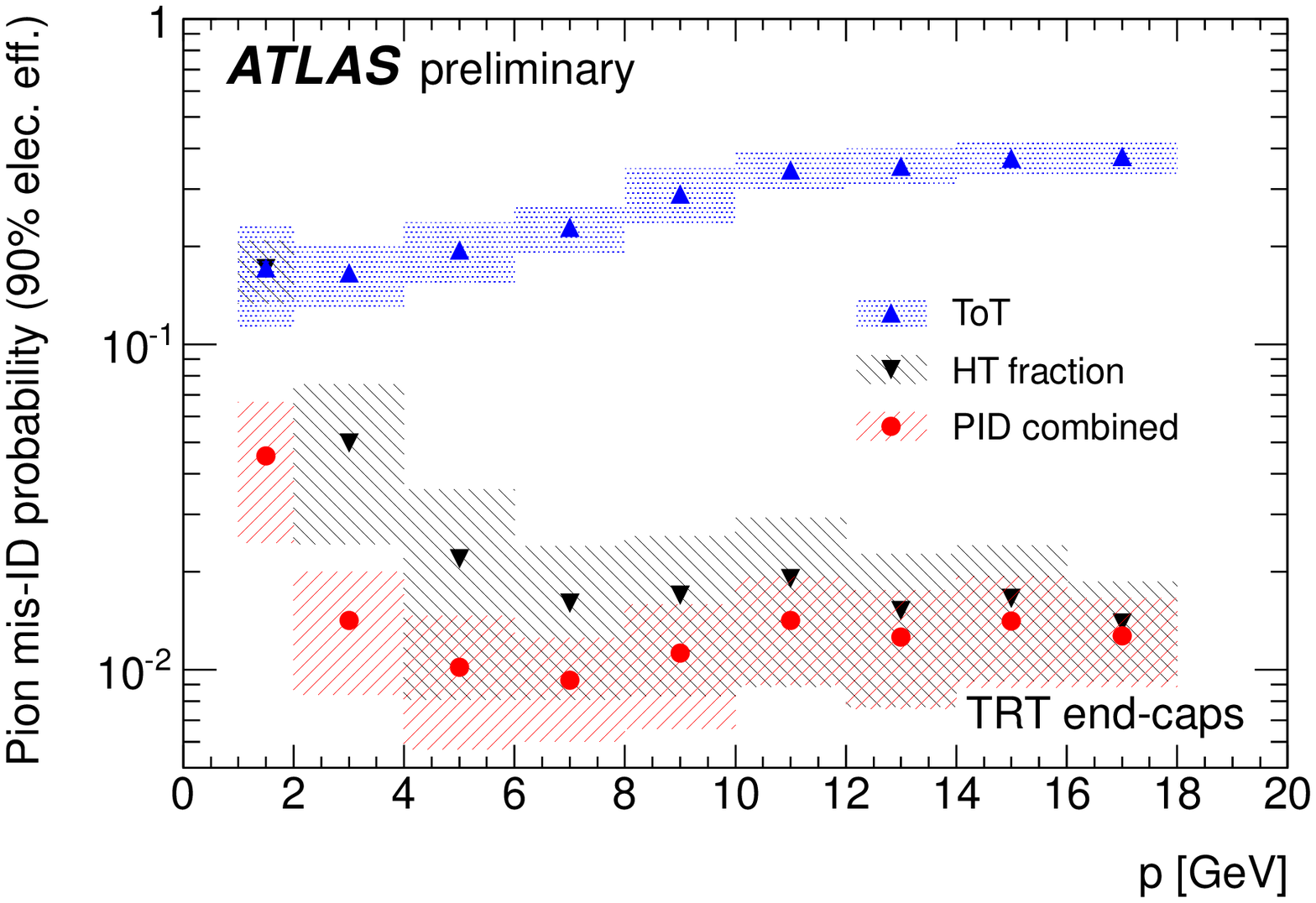}
    \label{fig:CombEndcap}}
  \caption{The pion mis-ID probability for selection criteria that give
    90\% electron efficiency, showing HT, ToT and their combination}
  \label{fig:Comb}
\end{figure}

The HT fraction and the ToT measurements can be combined to achieve
the best electron identification performance. To combine the HT and
ToT measurements, two likelihood functions are first formed based on
the discriminating variables: one for HT, and one for ToT. Since the
HT hits are not used for ToT discriminator, the two likelihoods are
assumed to be independent, and are multiplied to form a single
combined likelihood. The electrons are then selected by applying a cut
on the combined likelihood.  Again, a cut value that gives a 90\%
electron efficiency was determined in different momentum bins, and
applied to the pion sample to determine the efficiency for pions to
pass the same criterion. Figure \ref{fig:Comb} shows the pion
misidentification probability p$_{\pi\rightarrow e}$ at 90\% electron
efficiency as a function of momentum. The uncertainties are again
estimated by varying the selection criteria such that the electron
efficiency changed by $\pm$2\%. It should be noted that any
contamination of the pion sample with electrons above the TR threshold
will systematically bias the estimate of p$_{\pi\rightarrow e}$ by
roughly the same amount. As can be seen in the figure, the ToT-based
selection improves the pion rejection at p $<$ 10 GeV, where the
discriminating power of the HT is lower than at high momenta.

\section{Summary}

Studies in the early collision data collected with the ATLAS detector
have confirmed that electron identification based on transition
radiation measured by the TRT is performing well, and in some detector
regions even exceeds the performance obtained from the current
detector simulation. The pion misidentification probability for
selection criteria that give 90\% electron efficiency is about 5\%
(rejection factor 20) for the majority of the detector and as low as
1-2\% in the best performing detector regions. Analysis of data from a
dedicated run with different hardware settings confirmed that the
thresholds were close to their optimal value, and only small
adjustments were made in order to ensure stable performance under a
wide range of operating conditions. The transition radiation
measurement was used to identify electrons for the first W boson
production cross section measurement by ATLAS [17], as well as for the
W$^{+}$W$^{-}$ cross section measurement [18] and other analyses such as a
search for supersymmetry [19].  Time over threshold measurements
can be used to further improve the electron identification, in
particular for tracks with momentum less than 10 GeV. 


\bigskip 

\end{document}